\documentclass[twocolumn,showpacs,prb,amsmath,amssymb]{revtex4}
\usepackage{graphicx}
\usepackage{dcolumn}
\usepackage{bm}

\begin{document}
\title{Coupled dynamics of electrons and phonons in metallic nanotubes:\\
current saturation from hot phonons generation}
\author{Michele Lazzeri and Francesco Mauri}
\affiliation{Institut de Min\'eralogie et de Physique des
Milieux Condens\'es, 4 Place Jussieu, 75252, Paris cedex 05, France}
\date{\today}

\begin{abstract}
We show that the self-consistent dynamics of both phonons and
electrons is the necessary ingredient for the reliable description of
the hot phonons generation during electron transport in metallic
single-wall carbon nanotubes (SWNTs).  We solve the coupled Boltzmann
transport equations to determine in a consistent way the current
vs. voltage (IV) curve and the phonon occupation in metallic SWNTs
which are lying on a substrate.  We find a good agreement with
measured IV curves and we determine an optical phonon occupation which
corresponds to an effective temperature of several thousands K (hot
phonons), for the voltages typically used in experiments.  We show
that the high-bias resistivity strongly depends on the optical phonon
thermalization time. This implies that a drastic improvement of
metallic nanotubes performances can be achieved by increasing the
coupling of the optical phonons with a thermalization source.

\end{abstract}

\maketitle
\section{INTRODUCTION}
Much interest is currently devoted to the study of electronic
transport properties in carbon nanotubes (CNTs), both metallic and
semiconducting. The main reason is the possibility to use
CNTs in electronic integrated circuits thanks to catalizator-assisted
on-site growth~\cite{kong98,tseng04}.  Semiconducting CNTs are envisageables
as new components for transistors and logic circuits
~\cite{tans98,derycke01}.
On the other hand, metallic CNTs are particularly suited as a new type
of interconnects due to their small dimensions and the large electron
current density ($\sim$10$^9$ A/cm$^2$) they can support.  The current
vs. voltage (IV) curve of metallic single-wall carbon nanotube (SWNTs)
has been recently measured by several
groups~\cite{dekker00,javey04,park04}.  For voltages $\ge
0.2$~V, they observe a sudden increase of the resistivity which is due
to the scattering with optical phonons.  In long tubes, this leads to
a saturation current of $\sim$ 25~$\mu$A for voltages $\ge 5$~V. Such
behavior limits the performances of metallic SWNTs as
interconnects. The understanding of this phenomenon is, thus, a crucial
step toward finding methods to boost SWNTs performances and has
important technological consequences.

Two recent papers~\cite{lazzeri05,pop05} suggested the possibility
that, at high bias, the electron transport induces an anomalously high
optical-phonons occupation (hot phonons) which, in turn, induces an
increase of the resistivity. In Ref.~\onlinecite{lazzeri05}, this hypothesis
is formulated on the basis of the comparison between scattering
lengths (obtained from IV measurements
~\cite{dekker00,javey04,park04}) and electron-phonon coupling values
(obtained from ab-initio calculations~\cite{piscanec04} and inelastic
X-ray scattering measurements~\cite{maultzsch04}).  However, the
extent of Ref.~\onlinecite{lazzeri05} conclusions is limited, because
Ref.~\onlinecite{lazzeri05} does not provide neither a scheme to reproduce
experimental IV curves, nor a reliable quantitative determination of
the phonon occupation.  The conclusions of Ref.~\onlinecite{lazzeri05} apply
to tubes which are lying on a substrate. This is the very typical
situation encountered in experiments.

Ref.~\onlinecite{pop05} reports the comparison between IV curves measured on
SWNTs in two different situations: when the tubes are lying on a
substrate and when the tubes are suspended between the two electrodes.
Ref.~\onlinecite{pop05} concludes that hot phonons are present in the tubes
which are suspended but are absent in those which are lying on a
substrate (for bias $<1$~V).  The presence of hot phonons is clearly
demonstrated by the experimental observation of a negative
differential resistance in suspended tubes.  On the other hand, the
absence of hot-phonons is inferred from a simplified theoretical model
in which the electron scattering length (at zero temperature) is a
fitting parameter.  The claim of absence of hot phonons, for the tubes lying on
a substrate, is contradicting the conclusions of Ref.~\onlinecite{lazzeri05},
where the scattering length is obtained from ab-initio calculations of the
electron-phonon coupling.

Concluding, a quantitative model to describe the anomalous phonon heating
is not available. Moreover, given the contradicting conclusions of Refs.
~\onlinecite{lazzeri05,pop05}, it is not clear whether hot phonons generation
is present for tubes lying on a substrate. This last point is particularly
relevant. In fact, if the high-bias resistance measured for tubes lying on
a substrate~\cite{dekker00,javey04,park04} is due to hot phonons
~\cite{lazzeri05}, one could conceive some methods to reduce the optical
phonon temperature and thus diminish the high bias resistance.
This would not be possible if hot phonons were not present in tubes
lying on a substrate~\cite{pop05}.

In this paper, we solve the coupled Boltzmann transport equations for
both phonons and electrons to determine in a consistent way the IV
curve and the phonon occupation in metallic SWNTs. The quantitative
determination of the anomalous phonon heating during electron
transport allows us to settle the debate on the presence/absence of
hot phonons in nanotubes lying on a substrate.  We are interested in
the high bias ($> 0.2$~V) region, where the transport properties are
determined by the scattering with optical phonons.  We use the
zero-temperature electron scattering lengths obtained from ab-initio
calculations based on the density functional theory
~\cite{piscanec04,lazzeri05,DFPT}.

\section{DESCRIPTION OF THE MODEL}
\subsection{GENERAL CONSIDERATIONS}

In a metallic SWNT, there are only two phonons which can be generated
by transport electrons via a back-scattering process~\cite{lazzeri05}.
Using the graphene notation,
they correspond to the E$_{2g}$ LO (here LO stands for
longitudinal along the tube axis) mode at ${\bm \Gamma}$ and to the
A$'_1$ mode at {\bf K} (equivalent of the A$'_1$ at {\bf
K'})~\cite{nota_2phon}.  For simplicity, we label them ${\bm \Gamma}$
and ${\bf K}$.  $\tau_{bs}^{\bm \Gamma}$ and $\tau_{bs}^{\bf K}$ are
the corresponding electron scattering times, i.e. the average time an
electron can travel before emitting a phonon.  $\tau_{bs}=l_{bs}/v_F$,
where $l_{bs}$ is the scattering length and $v_F=8.39\times 10^7$~cm/s
is the Fermi velocity~\cite{piscanec04}.  According to precise
ab-initio calculations~\cite{lazzeri05}, at zero temperature
$\tau_{bs}^{\bm \Gamma}$ = 538 and $\tau_{bs}^{\bf K}=$
219~fs~\cite{nota01} (for tubes with a diameter of 2.0~nm, which is
the diameter typically used in experiments
~\cite{dekker00,javey04,park04}).

Electron transport can induce not-thermal high-occupation of phonons
if the generation time of the phonons ($\tau^{\bm \Gamma}_{bs}$ and
$\tau^{\bf K}_{bs}$) is shorter than their
thermalization time.  For metallic SWNTs, the optical-phonons
thermalization time $\tau_{th}$ is due to anharmonic scattering into
acoustic phonons ~\cite{nota_substr}.  Time-resolved terahertz
spectroscopy measurements~\cite{kampfrath05}, on graphite, reported an
optical-phonon thermalization time of 7~ps.  This value is comparable
with the optical-phonon linewidth of diamond (the other carbon
polymorph) $\sim$1~cm$^{-1}$, which corresponds to a relaxation time
of 5~ps, and is due to phonon-phonon scattering~\cite{lang}.  A
$\tau_{th}$ value of the same order is also obtained by preliminary
calculations of phonon-phonon scattering on graphite and nanotubes
~\cite{bonini} done with the ab-initio methods of
Refs.~\onlinecite{lang,shukla}.  Concluding, we expect $\tau_{th}$ to be of
the order of 5~ps.  Since $\tau^{\bm \Gamma}_{bs}$ and $\tau^{\bf
K}_{bs}$ are much smaller than $\tau_{th} \sim$5~ps, hot-phonon
generation is {\it a priori} expected to occur during electron
transport in metallic SWNTs.


\begin{figure}
\centerline{\includegraphics[width=85mm]{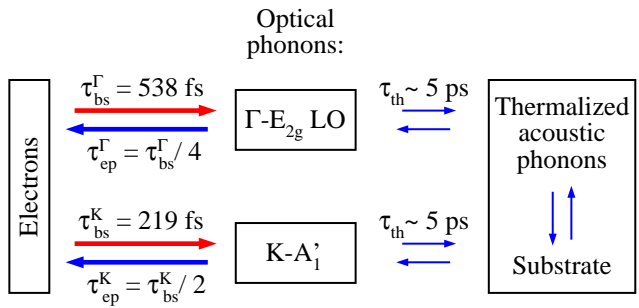}}
\caption{Scheme of the model used in this paper (see the text).}
\label{fig1}
\end{figure}
In the following, we consider the model schematized in Fig.~\ref{fig1}
Transport electrons can scatter with the optical phonons ${\bm
\Gamma}$, {\bf K} and {\bf K'} ({\bf K'} is equivalent of {\bf K} and is not
shown).  $\tau_{bs}^{\bm \Gamma}$ and $\tau_{bs}^{\bf K}$ are the
electron scattering times.  $\tau_{ep}^{\bm \Gamma}$ and
$\tau_{ep}^{\bf K}$ are phonon scattering times, i.e. the average time
a phonon lives before emitting an electron-hole pair.  For metallic
SWNTs, it can be shown~\cite{lazzeri-cond} that $\tau_{ep}^{\bm
\Gamma}=\tau_{bs}^{\bm \Gamma}/4$ and $\tau_{ep}^{\bf
K}=\tau_{bs}^{\bf K}/2$.  The optical phonons can, also, scatter into
acoustic phonons with a characteristic thermalization time
$\tau_{th}\sim 5$~ps. Both the acoustic phonons and the substrate are
thermalized at room temperature, and are acting as a thermal bath. The
present model is not valid if the tube is suspended between the two
electrodes, as in some of the experiments of Ref.~\onlinecite{pop05}.  In
this case, the acoustic phonons are not thermalized with the
environment and their occupation should be determined
self-consistently.  We are not considering this case.

\subsection{IMPLEMENTATION}

To study the evolution of the optical-phonons occupation during
electron transport, we consider the Boltzmann transport
equation.
In a metallic SWNT the electron gap is zero for two equivalent k-points
corresponding to {\bf K} and {\bf K'} in the graphite Brillouin zone. 
Two electron bands are crossing at the Fermi energy $\epsilon_F$
at both {\bf K} and {\bf K'}.
For the energy
range of interest ( $\pm$ 0.5~eV around $\epsilon_F$), the two bands
can be considered linear with slope $\pm\hbar |v_F|$, where $v_F$ 
is the Fermi velocity.
$f_{L/R}(k,x,t)$ is the probability to have an electron coming from the
left or right contact ($L$,$R$) with momentum $k$ ($k=0$ at the band crossing)
, at the position $x$.
States near {\bf K} or {\bf K'} have the same distribution $f$.
$f$ evolves in time ($t$) according to
\begin{equation}
[\partial_t \pm v_F\partial_x - e{\cal E}/\hbar\partial_k]f_{L/R}=
[\partial_t f_{L/R}]_c,
\label{eq1}
\end{equation}
where $v_F$ is the Fermi velocity and $e$ ($e>0$) is the electron charge.
For an applied voltage $V$, we consider a uniform electric field
${\cal E}=-V/L$, where $L$ is the tube length.
The collision term $[\partial_t f]_c$ describes the electron scattering.
The probability distribution is given by the stationary solution
of Eq.~\ref{eq1}. 
As explained in Ref.~\onlinecite{dekker00}, $[\partial_t f]_c$ can be considered
as the sum of three terms:
(1) the elastic scattering $[\partial_t f_L]_e = (v_F/l_e)(f_R-f_L)$,
where $l_e$ is the elastic scattering mean free path;
(2) the forward scattering from optical phonons $[\partial_t f_L]_{fs}$;
(3) the back scattering from optical phonons $[\partial_t f_L]_{bs}$.
Following Refs.~\onlinecite{park04,pop05} we assume $l_e=1600$~nm.
We neglect $[\partial_t f_L]_{fs}$ since it does not change the propagation
direction of electrons and, thus, should have a minor effect on the current.
Now, we consider an optical phonon $\nu={\bm \Gamma},{\bf K}$, with occupation
probability $n^\nu(k,x,t)$ (for the {\bf K} phonon, $k$ stands for K+$k$) .
$n$ evolves according to
\begin{equation}
[\partial_t + v^\nu_{ph}(k)\partial_x]n^\nu=[\partial_t n^\nu]_c,
\label{eq2}
\end{equation}
where $v^\nu_{ph}$ is the phonon velocity and $[\partial_t n^\nu]_c$ is the
collision term.
We use $v^{\bm \Gamma}_{ph}= {\rm sign}(k)~2.9\times 10^5$~cm/s and
$v^{\bf K}_{ph}= {\rm sign}(k)~7.2\times 10^5$~cm/s, from ab-initio
calculations~\cite{piscanec04}, where ${\rm sign}(k)$ is the sign of $k$.
We remark that, given the small phonon velocity (with respect to $v_F$)
the optical phonons do not have the time to diffuse along tubes
longer than 100~nm (the thermalization scattering lengths are
$l^{\bm \Gamma}_{th}=v^{\bm \Gamma}_{ph}\tau_{th}\sim 15$~nm and
$l^{\bf K}_{th}=v^{\bf K}_{ph}\tau_{th}\sim 36$~nm).
This was already pointed out in Ref.~\onlinecite{pop05}.
Thus, the resulting IV curve is only slightly affected by the
exact value of $|v_{ph}|$.
For given $x$ and $t$, 
\begin{widetext}
\begin{eqnarray}
\left[\partial_t f_L\right]_{bs} &=&\sum_\nu
\frac{1}{\tau^\nu_{bs}} \left\{
[1-f_L]f_R(-k^+)
-[1-f_R(-k^-)]f_L +
\right. \nonumber \\
&&\left.
[f_R(-k^+)-f_L]n^\nu(-k-k^+)
+[f_R(-k^-)-f_L]n^\nu(k+k^-)
\right\} \label{eq3}\\
\left[ \partial_t n^\nu \right]_c &=&\frac{1}{2\tau^\nu_{ep}} \left\{
\left[1-f_L\left(\frac{-k^+}{2}\right)\right]
f_R\left(\frac{k^-}{2}\right)+
\left[1-f_R\left(\frac{-k^-}{2}\right)\right]
f_L\left(\frac{k^+}{2}\right)+
\right. \nonumber \\
&&\left.
\left[
-f_L\left(\frac{-k^+}{2}\right)
+f_R\left(\frac{ k^-}{2}\right)
-f_R\left(\frac{-k^-}{2}\right)
+f_L\left(\frac{ k^+}{2}\right)
\right] n^\nu(k)
\right\} - \frac{1}{\tau_{th}}n^\nu(k),
\label{eq4}
\end{eqnarray}
\end{widetext}
where $k^\pm = k\pm(\omega^{\nu}/v_F)$, $f_{L}$ stands for $f_{L}(k)$,
the variables $x$ and $t$ are omitted, and $\nu={\bm \Gamma},{\bf K}$.
$\omega^{\nu}$ is the phonon pulsation,
$\hbar\omega^{\bm \Gamma} = 196.0$~meV, $\hbar\omega^{\bf K} = 161.2$~meV
and the $\omega^{\nu}$ dependence on $k$ can be neglected.
Notice that Eq.~\ref{eq4} can be obtained from
Eqs.~\ref{eq1},\ref{eq2}, and \ref{eq3},
imposing the conservation of energy and momentum in the back-scattering
processes.
We recall that, to be consistent with the Boltzmann treatment,
the scattering times $\tau^\nu_{bs}$ and $\tau^\nu_{ep}$ of
Eqs.~\ref{eq3} and~\ref{eq4} (Fig.~\ref{fig1})
are computed with zero phonon occupation~\cite{nota01}.

We impose the equilibrium distributions as boundary conditions:
\begin{eqnarray}
f_L(k,0)&=&f_R(-k,L)=\{{\rm exp} [\hbar v_F k/(k_B T)] +1\}^{-1} \nonumber \\
n^\nu(k>0,0)&=&n^\nu(k<0,L)=0 \nonumber,
\end{eqnarray}
where $T=300$~K and $k_B$ is the Boltzmann constant. 
The current is given by:
\[
I = \frac{4e}{h} \int (f_L-f_R) \hbar v_F~dk.
\]
The stationary solution of Eqs.~\ref{eq1} and ~\ref{eq2} is found by
numerical integration in time~\cite{nota_spline}.

The present approach is similar to that of
Refs.~\onlinecite{dekker00,javey04} but has two important
differences. First, in Refs.~\onlinecite{dekker00,javey04} the phonon
occupation is considered to be thermalized at room temperature
($n\simeq 0$).  Thus, in Refs.~\onlinecite{dekker00,javey04}
Eq.~\ref{eq4} is not taken into account and, in Eq.~\ref{eq3}, the
terms depending on $n$ are not considered.  Second, in
Refs.~\onlinecite{dekker00,javey04}, the scattering time $\tau_{bs}$
is considered as a parameter and its value is fitted to reproduce the
experimental IV curves, supposing that hot phonons are absent. In the
present work, we do not make any assumption on the phonon occupation
$n$ and we determine $n$ by solving the transport equation. Moreover,
the zero-temperature scattering times $\tau^\nu_{bs}$ are not fitted
to recover a better agreement with measurements, but are fixed to the
values obtained from ab-initio calculations
~\cite{piscanec04,lazzeri05,DFPT}.

\section{RESULTS}
\subsection{CURRENT VS. VOLTAGE}

\begin{figure}
\centerline{\includegraphics[width=75mm]{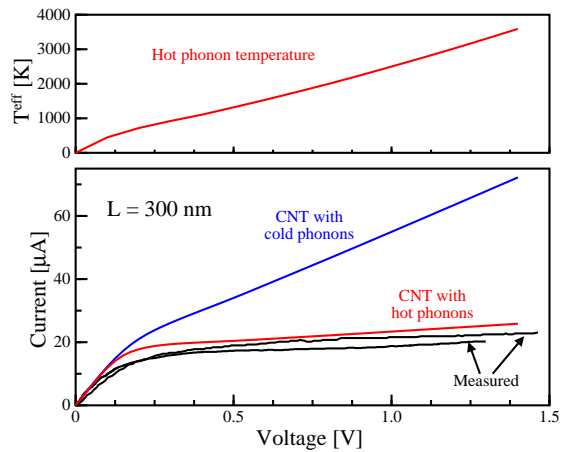}}
\caption{(Color online) Lower panel: current vs. voltage characteristic of a
300 nm long carbon nanotube (CNT). Calculations are done fixing the
phonon occupation to zero (cold phonons) or allowing
phonons to heat up (hot phonons, $\tau_{th}=5.31$~ps).
Measurements are from Refs.\protect\onlinecite{javey04,park04}.
Upper panel: phonon effective temperature of the Raman $G$ peak
(see the text) as a function of the voltage.
}
\label{fig2}
\end{figure}

In Fig.~\ref{fig2} we show the comparison between the measured IV
curves and our calculations done imposing the phonon occupation $n =
0$ (cold phonons) or allowing the $n$ to vary according to
Eqs.~\ref{eq2}, ~\ref{eq3} and ~\ref{eq4}, using $\tau_{th}=5.31$~ps
(hot phonons). This value corresponds to an anharmonic contribution
to the phonon linewidth (full width) $\gamma =$1~cm$^{-1}$, which is expected
from independent calculations and measurements
~\cite{kampfrath05,lang,bonini} (see discussion above).
{\it Imposing $n = 0$ the resulting IV curve dramatically underestimates
the experimental resistivity. On the contrary, if
the phonon occupation $n$ is determined self-consistently
the IV curve nicely reproduces the experimental data.}
Already at 0.2~V the hot-phonon and the cold-phonon curves significatively
differ. This implies that the hot
phonon generation dominates the resistivity behavior already
at V$>$0.2~V, also for tubes lying on a substrate.

\begin{figure}
\centerline{\includegraphics[width=75mm]{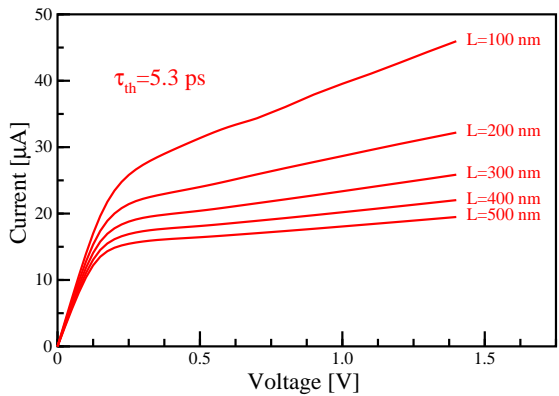}}
\caption{(Color online) Current vs. voltage curve calculated for various
tube lengths (L). $\tau_{th} = 5.31$~ps (see the text).}
\label{fig3}
\end{figure}

In Fig.~\ref{fig3} we show our calculations for tubes with different
lengths $L$. The dependence of the IV-curve on $L$ is similar to that
observed experimentally in Refs.~\onlinecite{javey04,park04}.
Despite the presence of the hot phonons, for bias $> 0.5$~V the
differential resistance $dV/dI$ is almost independent from the
voltage. 


\begin{figure}
\centerline{\includegraphics[width=85mm]{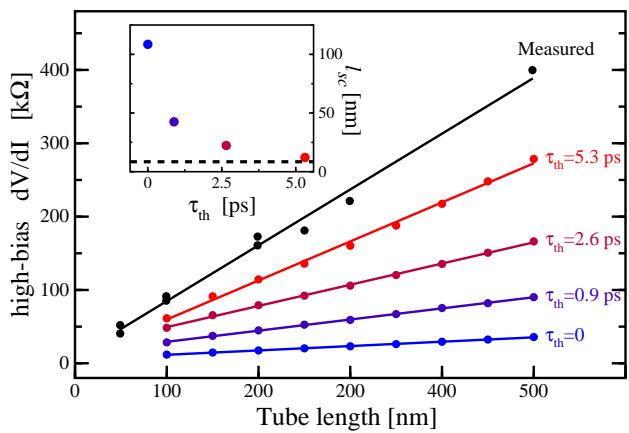}}
\caption{(Color online) High-bias resisitivity for various tubes
(see the text).
Measurements are from Ref.\protect\onlinecite{park04}.
Inset: the scattering length $l_{sc}$ (defined in Eq.~\protect\ref{eq5})
as a function of $\tau_{th}$. The dashed line correspond to the
experimental value $l_{sc}=8.5$~nm (see the text).
}
\label{fig4}
\end{figure}

We computed the $dV/dI$ at high bias (in between 1.0 and 1.4 V) for
tubes with different lengths $L$ (Fig.~\ref{fig4}).
Beside $\tau_{th}=$5.31 ($\gamma =1$~cm$^{-1}$),
we also considered $\tau_{th}=$2.65, 0.885, and 0~ps (corresponding to
$\gamma=$2,6,$+\infty$~cm$^{-1}$),
to investigate whether a significant reduction of the resistance can be
achieved by a reduction of $\tau_{th}$.
$dV/dI$ is clearly linear with
respect to $L$ and is, thus, possible to define a linear resistivity
$\rho$. For a one-dimentional channel with 4 sub-bands, in the
incoherent limit~\cite{datta} :
\begin{equation}
\rho=\frac{1}{L}\frac{dV}{dI} = \frac{1}{l_{sc}} \frac{h}{4e^2},
\label{eq5}
\end{equation}
where $h/(2e^2)=12.9~{\rm k}\Omega$ is the quantum of resistance and
$l_{sc}$ is the effective scattering length.  
 Eq.~\ref{eq5} can be used to obtain $l_{sc}$ from the measured resistivity
$\rho$.  Using this approach, several independent
experimental works~\cite{dekker00,javey04,park04,pop05} found a
scattering length $l_{sc}$ of the order of 10~nm, at high bias, for
tubes lying on a substrate.  We remark that $l_{sc}$ is not to be
confused with the $l_{bs}$ of the present paper. While $l_{bs}$ is
calculated for zero phonon occupation, $l_{sc}$ is an effective
scattering length that includes the effects of the finite
phonon occupation.  To compare our calculations with the measured $l_{sc}$, we
make a linear fit of the data in Fig.~\ref{fig4}, we obtain $\rho$ and,
thus, $l_{sc}$ through Eq.~\ref{eq5}. This is done for the different
$\tau_{th}$ values.  For $\tau_{th}=5.31$~ps (which is our best
$\tau_{th}$ estimate) we obtain $l_{sc}=$12~nm. Again, this compare
very well with $l_{sc}=$8.5~nm which we extracted from the
experimental values~\cite{park04} reported in Fig.~\ref{fig4}.

Given the fact that the IV curves strongly depend on the optical
phonons occupation (Fig.~\ref{fig2}), we expect that the thermalization
time $\tau_{th}$ might play a crucial role.
In fact, a smaller $\tau_{th}$ induces
a lower optical phonons occupation (Fig.~\ref{fig1}). In turn, this
results in lower resistivity or a higher scattering length
$l_{sc}$.  Indeed, for $\tau_{th}=$2.65, 0.885, and 0~ps, we obtain
$l_{sc}=$ 22, 42, and 108~nm, respectively (inset of Fig.~\ref{fig4}).
This strong dependence of $l_{sc}$ on $\tau_{th}$ implies that even a
small decrease of $\tau_{th}$ induces an important amelioration of
the nanotube performances. A higher optical phonon thermalization can be
obtained, e.g., by using a substrate characterized by vibrational
frequencies similar to those of a CNT~\cite{nota_substr}.

Finally, the calculations done with $\tau_{th}=0$ are
equivalent of imposing $n^\nu=0$ (cold phonons) in Eqs.~\ref{eq3},~\ref{eq4}.
For $n=0$, it is usually assumed that
$l_{sc}$ is a measure of the optical phonon back-scattering length,
i.e.  $(l_{sc})^{-1} = (l_{bs}^{\bm \Gamma})^{-1} + (l_{bs}^{\bf
K})^{-1}$.  Indeed, from our calculations for $n=0$ ($\tau_{th}=0$)
$l_{sc}= 108$~nm, which is 20\% smaller than $[(l_{bs}^{\bm
\Gamma})^{-1}+(l_{bs}^{\bf K})^{-1}]^{-1} = 131$~nm.

\subsection{PHONON OCCUPATION}

\begin{figure}
\centerline{\includegraphics[width=75mm]{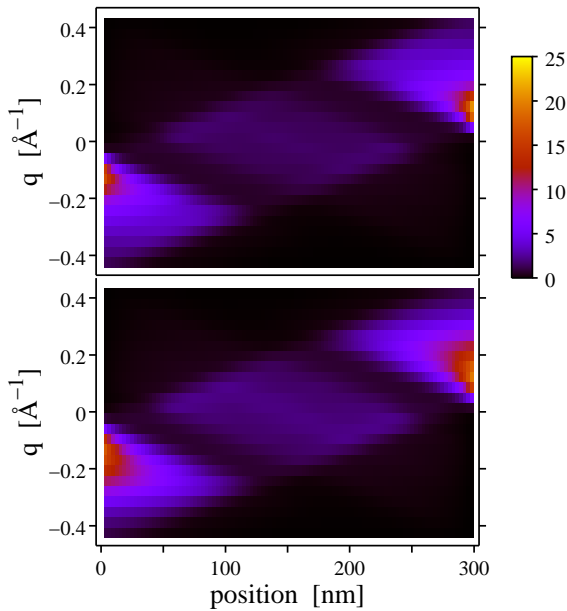}}
\caption{(Color online)
Phonon occupation $n$ as a function of the position along the tube,
$x$, and momentum $k$. Upper panel: ${\bm \Gamma}$-E$_{2g}$~LO phonon. Lower
panel: {\bf K}-$A'_1$. The tube is the same as in Fig.~\protect\ref{fig2},
with voltage 1.4 V.
}
\label{fig5}
\end{figure}

In Fig.~\ref{fig5}, we show an example of the computed phonon occupation
$n(k,x)$ at voltage 1.4~V, $L$=300~nm.
The phonon occupation is clearly distributed in a not uniform way.
Two important peaks with $n\sim20$ are present near the electrodes.
For a given phonon occupation $n(k,x)$, we define an effective temperature
$T^{eff}(k,x)$ as the corresponding temperature within the Bose-Einstein
statistics:
$n=1/\{exp[\hbar\omega/(k_BT^{eff})]-1\}$.
$n\sim20$ is extremely high and corresponds to
$T^{eff}\sim40000$~K.
Using simple arguments, it can be shown that $n(k,x)\ne 0$
only for $|k|<2Ve/(\hbar v_F)$.
This region is small compared to the Brillouin zone of
graphite~\cite{nota03}.  Thus, the hot phonons are indeed concentrated
around ${\bm \Gamma}$ and {\bf K}.
To evaluate the impact of the large phonon occupation on the nanotube
stability we compute
the average kinetic energy (per carbon atom) associated the hot-phonons
as
\[
E^{at} = \sum_\nu \frac{m^{\nu}\hbar\omega^{\nu}}{2\pi N^{at}}\int
n^\nu(k,x)~dk~dx,
\]
where $N^{at}$ is the number of atoms in the tube,
$m^{\bm \Gamma}=1$ and $m^{\bf K}=2$ (to take into account both
{\bf K} and {\bf K'}).
$E^{at}/k_B=50.6$~K,  at bias 1.4~V and $L$=300~nm.
Thus, the actual average kinetic energy per atom, associated to the hot
phonons,
is not particularly high despite the high values reached by the hot phonons
distribution, $n^{\nu}(k,x)$. 
Indeed, only a small subset of optical phonons is heated
in a limited region of the k-space.
Concluding, the presence of hot phonons having temperatures of several
thousands $K$ is compatible with the fact that the tubes are not melting
during electron transport.

As suggested in Ref.~\onlinecite{lazzeri05},
the occupation of the E$_{2g}$ phonon at $k=0$, $n^{\Gamma}(0,x)$,
is accessible experimentally using Raman spectroscopy. 
Indeed, this mode is responsible for the $G$ peak observed in Raman
spectroscopy~\cite{jorio02,oron05}.
The  Stokes and anti-Stokes $G$ peak intensities are
proportional to $n^{\Gamma}(0,x)+1$ and $n^{\Gamma}(0,x)$, respectively.
From the measured Stokes and anti-Stokes ratio one can extract
$n^{\Gamma}(0,x)$ of a SWNT during electron transport.
In the upper panel of Fig.~\ref{fig2} we show, as a function of
the applied bias, the effective temperature $T^{eff}$ of the Raman active
E$_{2g}$ phonon, obtained from the average over the tube length of
$n^{\bm \Gamma}(0,x)$.
Already at 1.0 V, we find a $T^{eff}= 2500~K$
(Fig.~\ref{fig2}), corresponding to $n=$~0.7.

Finally, Refs.~\onlinecite{lazzeri05,pop05} built two oversimplified models to
describe the hot-phonon generation. Both models are based on
the assumption that there is a simple relation between the effective
scattering length $l_{sc}$, the zero-temperature back-scattering length
$l_{bs}$, and the average phonon occupation $n$:
\begin{equation}
l_{sc} = \frac{l_{bs}}{2n+1}
\label{eq7}
\end{equation}
(see Eq.10 of Ref.~\onlinecite{lazzeri05} and lines 4-12, page 2 of Ref.
~\onlinecite{pop05}).
Alhtough a priori reasonable, Eq.~\ref{eq7} is in contradiction with
the observation that the differential resistance (and hence $l_{sc}$)
is constant with the voltage at high-bias (Fig.~\ref{fig3}), while the
phonon occupation $n$ is increasing with the voltage (upper panel of
Fig.~\ref{fig2}).  Therefore Eq.~\ref{eq7} can not be used to
construct quantitative models for the description of hot phonons in
metallic SWCNs.  The high-bias linearity of the IV curve, that is
observed experimentally in tubes with $L<500$~nm, might have led
Refs.~\onlinecite{dekker00,javey04,park04,pop05} into neglecting hot-phonons
effects for tubes lying on a substrate.

\section{CONCLUSIONS}

In conclusion, the explicit and self-consistent dynamics of both of
phonons and electrons is the necessary ingredient for the reliable
description of the hot phonons generation during electron transport in
metallic SWNTs.  We find a remarkable agreement with the experimental
IV curves.  We have shown that the hot phonon generation can not be
properly described using the simplified models reported in
Refs.~\onlinecite{lazzeri05,pop05}.  We demonstrate the presence of hot
phonons for voltages $>0.2$~V also for tubes which are lying on a
substrate.  We have shown that the high-bias resistance strongly
depends on the thermalization time of the optical phonons.  As a
direct consequence, an important improvement of metallic nanotubes
performances is, in principle, possible by increasing the
optical-phonons thermalization time.

\section*{ACKNOWLEDGMENTS}

We thank N. Bonini, N. Marzari and G. Galli for useful discussions.
Calculations were performed at IDRIS (Orsay, France); project 051202.



\begin{thebibliography}{99}


\bibitem{kong98}
J. Kong, H.T. Soh, A.M. Cassel, C.F. Quate, and H. Dai,
Nature (London) {\bf 395}, 29 (1998).

\bibitem{tseng04}
Y.-C. Tseng, P. Xuan, A. Javey,  R. Malloy, Q. Wang, J. Bokor, and H. Dai,
Nano Lett. {\bf 4}, 123 (2004).


\bibitem{tans98}
S. Tans, A. Verschueren, and C. Dekker,
Nature (London) {\bf 393}, 49 (1998).

\bibitem{derycke01}
V. Derycke, R. Martel, J. Appenzeller, and Ph. Avouris,
Nano Lett. {\bf 1}, 453 (2001).

\bibitem{dekker00}
Z. Yao, C.L. Kane, and C. Dekker,
Phys. Rev. Lett. {\bf 84}, 2941 (2000).

\bibitem{javey04}
A. Javey, J. Guo, M. Paulsson, Q. Wang, D. Mann, M. Lundstrom, and H. Dai,
Phys. Rev. Lett. {\bf 92}, 106804 (2004).

\bibitem{park04}
J-Y. Park, S. Rosenblatt, Y. Yaish, V. Sazonova, H. \"Ust\"unel, S. Braig,
T.A. Arias, P.W. Brouwer, and P.L. McEuen,
Nano Lett. {\bf 4}, 517 (2004).

\bibitem{lazzeri05}
M. Lazzeri, S. Piscanec, F. Mauri, A.C. Ferrari, and J. Robertson,
Phys. Rev. Lett. {\bf 95}, 236802 (2005).

\bibitem{pop05}
E. Pop, D. Mann, J. Cao, Q. Wang, K. Goodson, and H. Dai,
Phys. Rev. Lett. {\bf 95}, 155505 (2005).

\bibitem{piscanec04}
S. Piscanec, M. Lazzeri, F. Mauri, A.C. Ferrari, and J. Robertson,
Phys. Rev. Lett. {\bf 93}, 185503 (2004).

\bibitem{maultzsch04}
J. Maultzsch, S. Reich, C. Thomsen, H. Requardt, and P. Ordejon,
Phys. Rev. Lett. {\bf 92}, 075501 (2004).


\bibitem{DFPT}
S. Baroni, S. de Gironcoli, A. Dal Corso, and P. Giannozzi,
Rev. Mod. Phys. {\bf 73}, 515 (2001).

\bibitem{nota_2phon}
Because of the momentum conservation, only the phonons near ${\bm
\Gamma}$ or ${\bf K}$ can be involved in a scattering process with
electrons near the Fermi level. Moreover, the ${\bm \Gamma}$-E$_{2g}$
LO mode and the ${\bf K}$-A$'_1$ are the only phonons with a not
negligible electron-phonon coupling~\protect\cite{lazzeri05} for
back-scattering events. 

\bibitem{nota01}
Following Ref.~\protect\onlinecite{lazzeri05},
the scattering lengths are proportional to the tube diameter $d$,
in particular $l_{bs}^{\bm \Gamma} = 225.5~d$
for the ${\bm \Gamma}$-E$_{2g}$ LO mode and
$l_{bs}^{\bf K} = 91.9~d$ and  for the ${\bf K}$-A$'_1$.
These values are computed using the Fermi golden rule,
with zero phonon occupation.
$\tau^{\bm \Gamma}_{bs}=l^{\bm \Gamma}_{bs}/v_F$ and
$\tau^{\bf K}_{bs}=l^{\bf K}_{bs}/v_F$.

\bibitem{nota_substr}
In general, $\tau_{th}$ is also determined by the direct coupling of
the SWNT optical phonons with the substrate.  Such coupling is
relevant when the optical phonons frequencies are similar to those of
the substrate.  In the typical experimental situation, we expect this
coupling to be negligible because the nanotube is lying on a SiO$_2$
substrate.  SiO$_2$ phonon frequency are smaller than the SWNTs
optical phonons frequencies.

\bibitem{kampfrath05}
T. Kampfrath, L. Perfetti, F. Schapper, C. Frischkorn, and M. Wolf,
Phys. Rev. Lett. {\bf 95}, 187403 (2005).

\bibitem{lang}
G.\ Lang, K.\ Karch, M.\ Schmitt, P.\ Pavone, A.P.\ Mayer, R.K.\ Wehner,
and  D.\ Strauch,
Phys. Rev. B {\bf 59 }, 6182 (1999).

\bibitem{bonini}
N. Bonini, M. Lazzeri, F. Mauri, and N. Marzari{\it unpublished}.

\bibitem{shukla}
A. Shukla, M. Calandra, M. d'Astuto, M. Lazzeri, F. Mauri,
C. Bellin, M. Krisch, J. Karpinski, S.M. Kazakov, J. Jun, D. Daghero,
and K. Parlinski,
Phys. Rev. Lett. {\bf 90}, 095506 (2003).

\bibitem{lazzeri-cond}
This result can be easily derived considering the life times
(of both electrons and phonons) as given by the Fermi golden rule. See:
M. Lazzeri, S. Piscanec, F. Mauri, A.C. Ferrari, and J. Robertson,
arXiv.org/cond-mat/0508700.

\bibitem{nota_spline}
The probabilities $f(k,x)$ and $n(k,x)$ are defined on
a grid of points equally spaced by $\Delta k = 0.01~{\rm eV}/(\hbar
v_F)$ and $\Delta x$ going from 0.2 to 1 nm, depending on the tube
length.  $f$ was evolved in time according to Eq.~\protect\ref{eq1} using a
finite time step of $\Delta x/v_F$. The values of $f$ and $n$,
that are necessary for the next time step, are obtained
via a third-order polynomial spline interpolation in the $k$ coordinate.

\bibitem{datta}
S.Datta, {\it Electronic Transport in Mesoscopic Systems},
Cambridge University Press (London, 1995).


\bibitem{nota03}
For the voltage $V=1.4$~V (used in Fig.~\protect\ref{fig2})
$2Ve/(\hbar v_F)=0.51$~\AA$^{-1}$, which is to be compared with
$2\pi/a=2.55$\AA$^{-1}$, where $a$ is the graphite lattice parameter.

\bibitem{jorio02}
A. Jorio, A. G. Souza Filho, G. Dresselhaus, M. S. Dresselhaus, A. K. Swan,
M. S. \"Unl\"u, B. Goldberg, M. A. Pimenta, J. H. Hafner, C. M. Lieber,
and R. Saito,
Phys. Rev. B {\bf 65}, 155412 (2002).

\bibitem{oron05}
M. Oron-Carl, F. Hennrich, M. M. Kappes, H. v. Lohneysen, and R. Krupke,
Nano Lett. {\bf 5}, 1761 (2005).

\end{thebibliography}
\end{document}